\documentclass[aps,pre,twocolumn,showpacs,superscriptaddress,preprintnumbers,amsmath,amssymb]{revtex4}
\usepackage{amsmath}
\usepackage{graphicx}
\usepackage{dcolumn}
\usepackage{bm}
\usepackage{float}

\begin{document}

\title{Magnetic instability in a dilute circular rarefaction wave}


\author{M. E. Dieckmann}\thanks{Electronic mail: Mark.E.Dieckmann@itn.liu.se}
\affiliation{Department of Science and Technology (ITN), Linkoping University,
60174 Norrkoping, Sweden}

\author{G. Sarri}
\affiliation{Centre for Plasma Physics, School of Mathematics and Physics,  
Queen's University of Belfast, Belfast BT7 1NN, United Kingdom}

\author{M. Borghesi}
\affiliation{Centre for Plasma Physics, School of Mathematics and Physics,  
Queen's University of Belfast, Belfast BT7 1NN, United Kingdom}


\date{\today}


\begin{abstract}
The generation of a magnetic field in a circular rarefaction wave is examined in form of
a 2D particle-in-cell (PIC) simulation. Electrons with a temperature of 32 keV are uniformly
distributed within a cloud with a radius of 14.2 electron skin depths. They expand under
their thermal pressure and carry with them the cold protons, which are initially concentrated 
in a hollow ring at the boundary of the electron cloud. The interior of the ring contains an
immobile positive charge background that compensates for the electron charge. The protons
expand in form of a circularly symmetric rarefaction wave and they extract energy from the
electrons. A thermal anisotropy of the electrons develops and triggers through a Weibel-type
instability the growth of TM waves within the plasma cloud, which acts as a waveguide. The 
changing cross section of this waveguide introduces a coupling between the TM wave and a
TE wave and in-plane magnetic fields grow. The relevance of the simulation results to a 
previous experimental study of a laser-ablated wire is discussed.   
\end{abstract}
\maketitle

\section{Introduction}

Magnetic instabilities in expanding plasmas or in density gradients are of interest for 
laser fusion, where magnetic fields can reduce the particle's mobility \cite{Inertial}, 
and for astrophysical plasmas, where they can generate magnetic fields from noise levels. 
Magnetic field generation in the interstellar medium (ISM) \cite{Interstellar} by the 
interplay of the electrons of the hot ionized medium with the spatially nonuniform (ISM) 
plasma or the amplification of magnetic fields in the turbulent plasma close to supernova 
remnant (SNR) shocks \cite{ShockMag,Bell,Pohl} are examples. 

The growth of magnetic fields in rarefaction waves on an electron timescale
\cite{Grismayer,Thaury,Sarri1,Sarri2} has recently been observed experimentally \cite{WireExp} 
and with particle-in-cell (PIC) simulations \cite{Thaury,WireSim}. The instability is driven
by a thermal anisotropy and is thus similar to the Weibel instability 
\cite{Weibel1,Weibel2,Palodhi,Stockem1,Stockem2}. It is the result of the electron's slowdown 
by the ambipolar electrostatic field, which is sustained by the plasma density gradient of 
the rarefaction wave. This electrostatic field counteracts the charge separation along the 
plasma density gradient that arises from the difference in the thermal speeds of electrons 
and ions. It accelerates the ions and slows down the electrons along this direction, which 
generates the thermal anisotropy.


Previous simulations \cite{Grismayer,Thaury,WireSim} have considered systems, in which hot 
electrons accelerate an equal number of initially cold ions. Here the acceleration of protons, 
which are distributed in form of a hollow ring, by the ambipolar electrostatic field is 
examined with a PIC simulation. This proton ring distribution is a good approximation of the 
cross section of the ions of a laser-heated wire, as long as it is located far away from the 
laser impact point. 

The ablation of the wire and the resulting magnetic instabilities have been examined 
experimentally in Ref. \cite{WireExp}. The wire in the experiment is composed of heavy ions, 
which can not be accelerated to high speeds by the expanding electrons. The light protons 
that feed the rarefaction wave originate from surface impurities and they are approximated 
here by the hollow ring distribution. The reduced number of mobile protons  
dilutes the rarefaction wave. This is also observed in the experiment where its number density 
is $\sim 10^{18}\textrm{cm}^{-3}$ \cite{WireExp}, which is well below the solid ion number 
density of the wire. However, only the proton distribution in our simulation is realistic with 
respect to the experiment. The electrons have the temperature 32 keV and are spatially uniform 
within the plasma cloud, while the electrons in the experiment have MeV energies and are 
confined to the wire's surface \cite{Surface}. Reducing the electron 
temperature and distributing the electrons over a wide interval is computationally efficient 
and it ensures that the overall thermal energies in the experiment and simulation are comparable.

The purpose of our simulation is threefold. We want to determine if the gradient-driven 
magnetic instability always develops in the rarefaction wave or if competing instabilities 
can outrun it. Secondly, we want to determine if the lower proton density results in a weaker 
thermal anisotropy of the electrons and, thirdly, if and how the expansion of the dilute 
rarefaction wave differs from the dense one in Ref. \cite{WireSim}. 

Our results are as follows. The protons at the front of the dilute rarefaction wave are 
accelerated to about the same speed within the same time interval as those in the dense one 
in Ref. \cite{WireSim}. This confirms our expectation. The electrostatic potential of the 
plasma with respect to the surrounding vacuum is fully determined by the thermal pressure 
of the electrons, which is the same here and in our simulation in Ref. \cite{WireSim}. The
stronger reduction of the electron's thermal energy in Ref. \cite{WireSim} compared to the one
here does not lead to detectable differences in the plasma expansion. The same magnitude of 
the electrostatic potential and the therefrom resulting equal deceleration of the electrons 
in the direction of the density gradient imply that the electron's thermal anisotropies here 
and in Ref. \cite{WireSim} are equally strong. However, the much lower electron number density 
in the rarefaction wave we model here delays the onset of the gradient-driven magnetic 
instability. It develops instead in the dense core of the plasma cloud and the strong magnetic 
fields diffuse out into the rarefaction wave. 

The magnetic fields grow in the radial interval of the plasma cloud in which the temperature 
anisotropy and the number density of the electrons are large. The magnetic field source is 
thus a Weibel-type instability and not the competing thermoelectric instability \cite{Tidman}. 
The latter is inefficient in our case study, because the electron density gradient is aligned 
with the electron temperature gradient and because the expanding rarefaction wave is circularly 
symmetric. 

We also observe here the same secondary magnetic instability as in Ref. \cite{WireSim}, which 
yields the growth of in-plane magnetic fields. Our present simulation setup confines the secondary
wave in the plasma cloud's core, while it developed in the expanding rarefaction wave in Ref. 
\cite{WireSim}. This confinement simplifies the interpretation of the data. Our results suggest 
that the in-plane magnetic fields are generated by a mode conversion of a TM wave into a TE 
wave. This is a well-known process in waveguides with a variable cross section, which are important
in antenna theory \cite{Waveguide}. The variable cross section is here the result of the proton 
expansion.  

The structure of this paper is as follows. The PIC simulation scheme and the initial conditions
are summarized in Section 2. Section 3 presents the simulation results, which are discussed in
Section 4.

\section{The simulation code and the initial conditions}

A particle-in-cell (PIC) code approximates a plasma by an ensemble of computational particles 
(CPs) and it uses their collective charge distribution $\rho (\mathbf{x})$ and current 
distribution $\mathbf{J}(\mathbf{x})$ to evolve in time the electromagnetic fields on a 
spatial grid. The electric $\mathbf{E}$ and magnetic $\mathbf{B}$ fields update in turn the 
momentum of each CP through the relativistic Lorentz force equation. The PIC scheme is 
discussed in detail in Ref. \cite{Dawson}. 

Most codes evolve the electromagnetic fields through the discretized forms of the Amp\'ere's 
and Faraday's laws
\begin{eqnarray}
\frac{\partial \mathbf{E}}{\partial t} = \frac{1}{(\mu_0 \epsilon_0)}\nabla \times \mathbf{B} - 
\frac{1}{\epsilon_0} \mathbf{J},\\
\frac{\partial \mathbf{B}}{\partial t}=-\nabla \times \mathbf{E}
\end{eqnarray}
and they fullfill Gauss' law $\nabla \cdot \mathbf{E} = \rho / \epsilon_0$ and $\nabla \cdot 
\mathbf{B}=0$ either as constraints or through correction steps. The relativistic Lorentz
force equation 
\begin{equation}
\frac{d\mathbf{p}_j}{dt} = q_i \left [ \mathbf{E}(\mathbf{x}_j) + \mathbf{v}_j \times 
\mathbf{B}(\mathbf{x}_j) \right ]
\end{equation}
is used to update the momentum $\mathbf{p}_j$ of the $j^{th}$ particle of species $i$. The
collective behavior of the ensemble of the CPs of species $i$ approximates well that of a plasma 
species, provided that the charge-to-mass ratio of the plasma particles equals the ratio $q_i / 
m_i$ of the CPs, that the plasma is collision-less and that the statistical representation of the 
computational plasma is adequate. We use the numerical scheme discussed in Ref. \cite{Eastwood}.

\begin{figure}\suppressfloats
\includegraphics[width=\columnwidth]{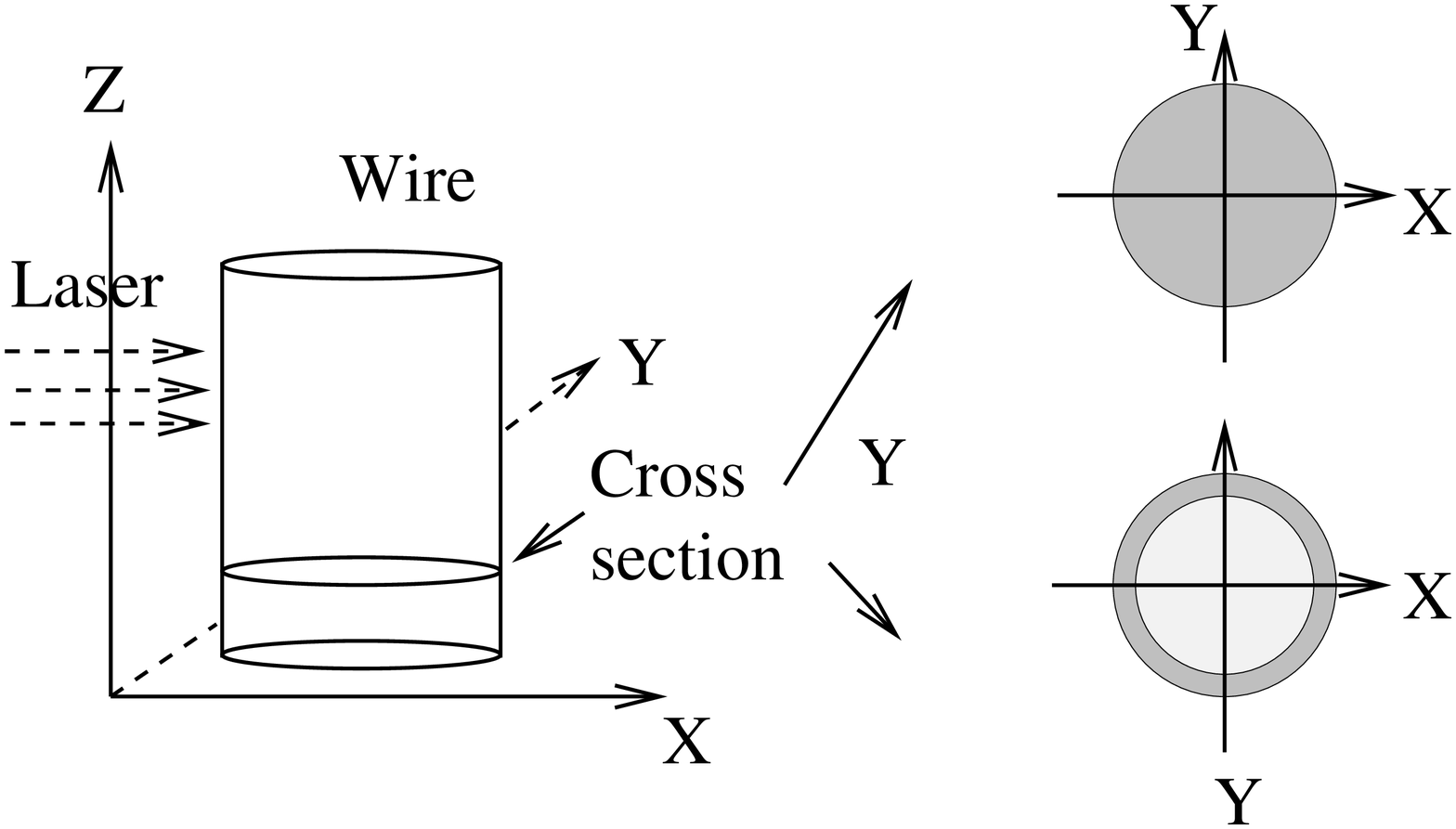}%
\caption{The initial conditions: The experimental setup is shown to the left, where we assume that 
a long wire is aligned with $z$. The initial electron distribution in the simulation is shown to 
the upper right and the proton ring distribution (dark gray shade) and the positively charged 
immobile background (light gray shade) to the lower right.}\label{f1}
\end{figure}

Our initial conditions and their connection to the experiment are visualized in Fig. \ref{f1}.
The axis of the wire on the left hand side is parallel to $\mathbf{z}$. The simulation resolves 
a cross-section of this wire in the x-y plane with an origin $x=0,y=0$ in the center of the wire. 
This cross-section has a z-coordinate that is sufficiently far away from that of the laser impact 
point so that we do not have to model the laser pulse in the simulation. 

In the experiment, the hot electrons stream uniformly from the laser impact point along the wire's 
surface \cite{Surface} to the cross section that corresponds to our simulation plane. We thus 
approximate the wire's cross section by the circular plasma cloud shown on the right hand side 
of Fig. \ref{f1}. Hot electrons fill the entire cross section of the plasma cloud with radius 
$r=r_W$. Their spatially uniform number density $n_0$ within the cloud gives the plasma frequency 
$\omega_p$. The mobile protons fill a hollow ring with the outer radius $r=r_W$ and with the inner 
radius $r=0.95r_W$. The interior $r<0.95r_W$ of the hollow ring contains an immobile positive charge 
background. The charge density of the electrons equals at any location $x,y$ with $x^2+y^2 < r^2$ 
that of the positive charge carriers and the mean speeds of both mobile species are zero. No net 
charge and no net current are thus initially present. All initial electromagnetic fields are thus 
set to zero. The cloud is immersed in a vacuum and the boundary conditions of the simulation box 
are periodic.

The outer cloud radius $r_W = 14.2 \lambda_e$, where $\lambda_e = c/\omega_{p}$ is the electron 
skin depth within the cloud. The temperature of the relativistic Maxwellian distribution that 
represents the electrons is 32 keV and their thermal speed is $v_e = c/4$. The bulk of the electrons 
thus moves at nonrelativistic speeds, which allows us to easily decompose their radial and azimuthal 
velocity components. The electron Debye length $\lambda_d = v_e/\omega_p$ equals $5.6 \Delta_x$, 
where $\Delta_x$ is the side length of a quadratic grid cell. We use the correct mass ratio between 
the electrons and the protons in the ring distribution. The proton temperature is 10 eV and their 
thermal speed $v_p = 3.1 \times 10^4$ m/s. 

The quadratic area $L \times L$ of the 2D simulation box with the side length $L=75.5\lambda_e$ is 
resolved by $N_g = 1700$ grid cells in the $x$ and $y$ directions. We represent the electrons by 
$N_e = 8 \times 10^8$ CPs and the protons by $N_p = N_e$ particles. Since the protons occupy a 
smaller volume, their numerical weight is lower. A time interval $t\omega_p = 800$ is resolved. We 
normalize time to $\omega_p$, space to the initial cloud radius $r_W$ and the electron and proton 
velocities to their respective initial thermal speeds $v_e$ and $v_p$. The electric and magnetic 
fields are normalized to $m_e \omega_p c /e$ and $m_e \omega_p / e$, respectively.

\section{Simulation results}

The kinetic energy of electrons with the mass $m_1$ is $K_1 (t) = m_1 c^2 \sum_{j=1}^{N_e} \left [ 
\Gamma_j - 1 \right ]$, where the summation is over all computational electrons with the Lorentz 
factors $\Gamma_j$ and $K_0 \equiv K_1 (0)$. The kinetic energy of the mobile protons with mass 
$m_2=1836 \, m_1$ is $K_2 (t) = m_2 c^2 \sum_{j=1}^{N_p} \left [ \Gamma_j - 1 \right ]$. The energy 
of the in-plane electric field is $E_{E\perp}(t) = (\Delta_x^3 / 2\epsilon_0) \sum_{i,j=1}^{N_g} 
E_p^2(i.j.t)$, where $E_p (i,j,t) = {[E_x^2(i,j,t) + E_y^2 (i,j,t)]}^{1/2}$. The energy density of 
the out-of-plane magnetic field $E_{Bz}(t) = (\Delta_x^3/2\mu_0)\sum_{i,j=1}^{N_g} B_z^2(i.j.t)$. 

\begin{figure}\suppressfloats
\includegraphics[width=\columnwidth]{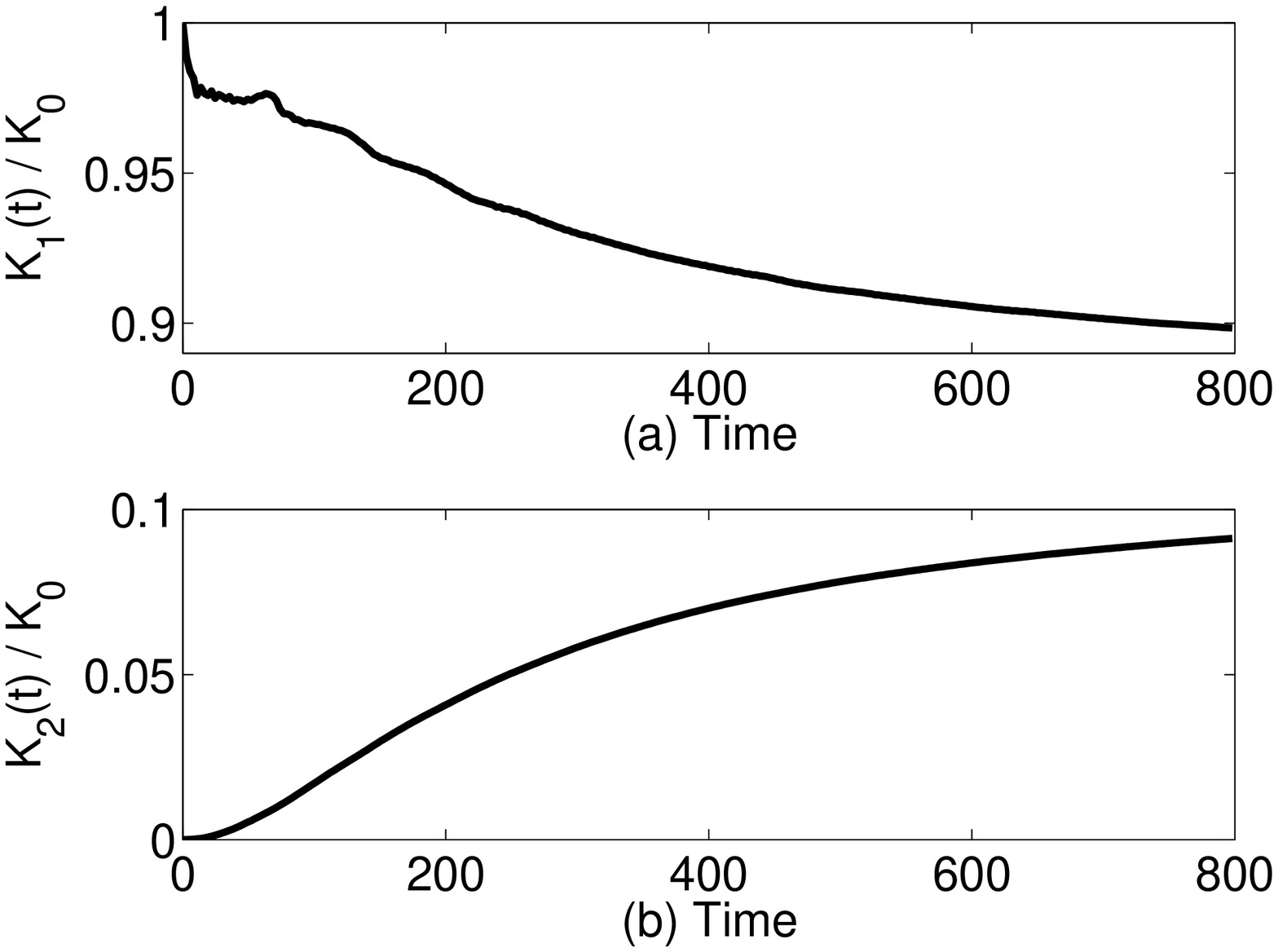}%

\includegraphics[width=\columnwidth]{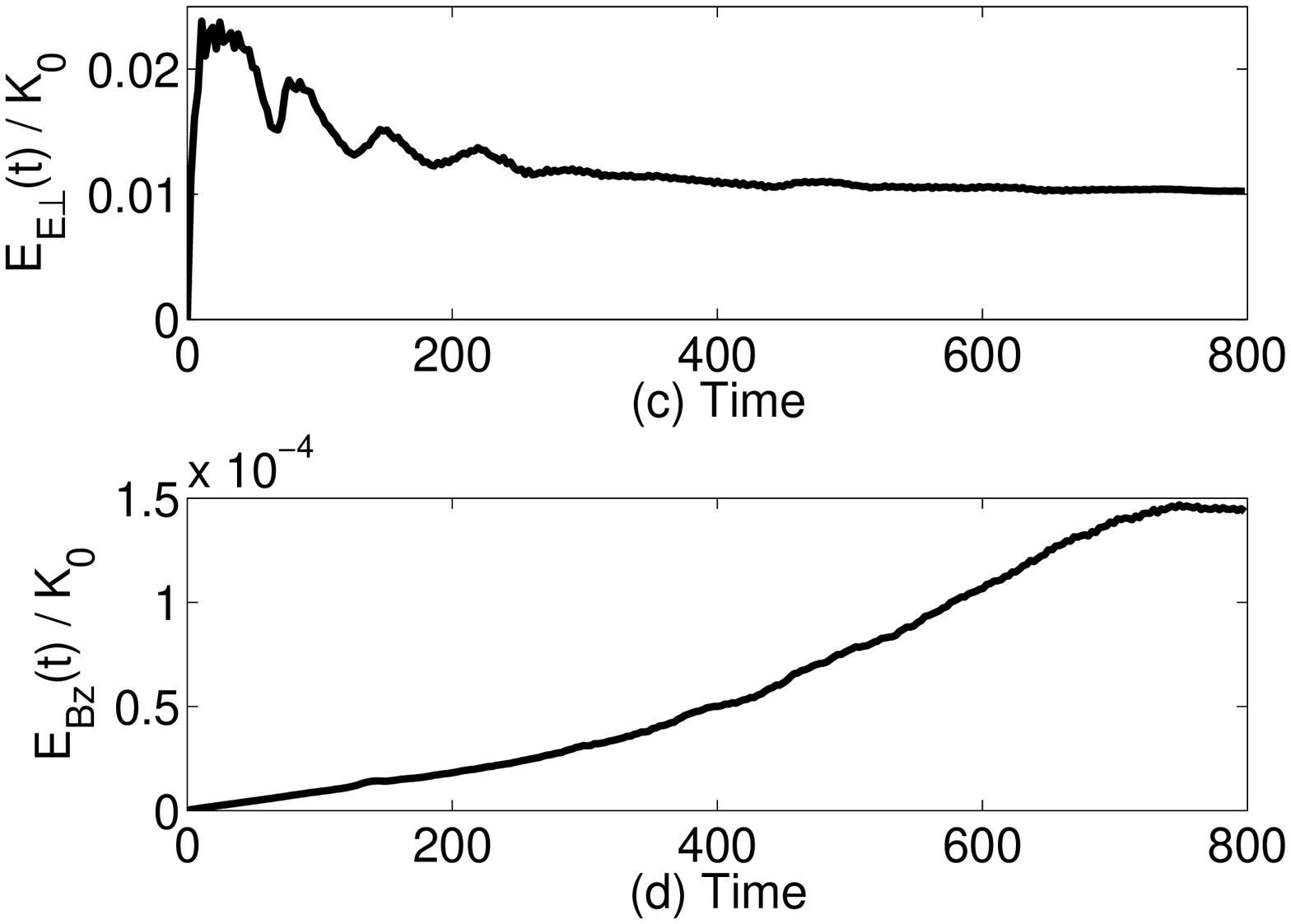}%
\caption{Energy densities in units of the initial electron thermal energy: The 
electron energy (a), the proton energy (b), the energy density of the in-plane 
electric field (c) and of the out-of-plane magnetic field (d).}\label{f2} 
\end{figure}

Figure \ref{f2} shows their time evolution.
The electrons sustain a rapid energy loss during $0<t<20$. This energy is transfered to the 
ambipolar electric field, which grows and saturates during this time. This field accelerates 
the protons and the electrons have transfered about 10\% of their initial energy to them at 
$t=800$. The initial oscillations of $E_{E\perp}$ have damped out at $t\approx 200$ and 
$E_{E\perp}$ reaches a steady state value of $\approx 10^{-2}K_0$. $E_{Bz}$ grows initially 
slowly. The faster growth of the magnetic energy in the time interval $400 < t < 700$ is followed 
by its saturation. The magnetic energy remains well below that found in Ref. \cite{WireSim} and 
about two orders of magnitude below the electric one. We will examine now in more detail the field 
and particle distributions at the time $t=27$ when the electrostatic field reaches its 
peak value, at $t=500$ when the magnetic field grows fastest and at $t=800$ when the magnetic 
field saturates.

\begin{figure}\suppressfloats
\includegraphics[width=\columnwidth]{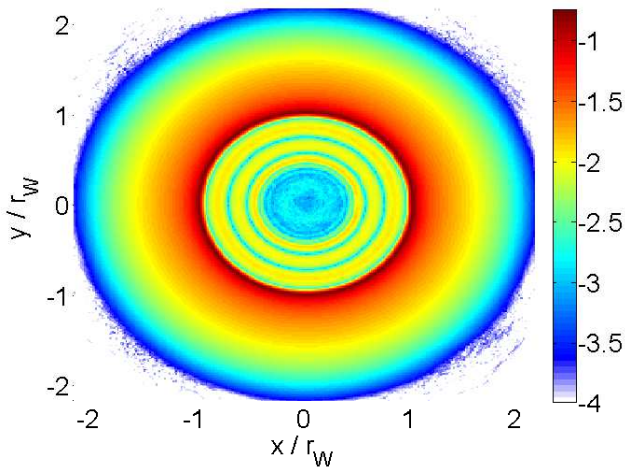}%
\caption{(Color online) The 10-logarithmic modulus of the electric field $|E_p(x,y)|$ 
sampled at the time $t=27$ (Enhanced online).}\label{f3}
\end{figure}

The modulus of the in-plane electric field at $t=27$ is shown in Fig. \ref{f3}. The electric
field has the expected circular symmetry. It peaks at $r\approx r_W$. It gradually decreases for 
increasing values of $r$ and reaches noise levels at $r\approx 2 r_W$. Circular electric field 
oscillations are visible in the cloud's core $r<r_W$. Since the positive charged background is 
immobile in this region, these electrostatic waves must be Langmuir waves. 
The magnetic field remains at noise levels at this time (not shown).

\begin{figure}\suppressfloats
\includegraphics[width=\columnwidth]{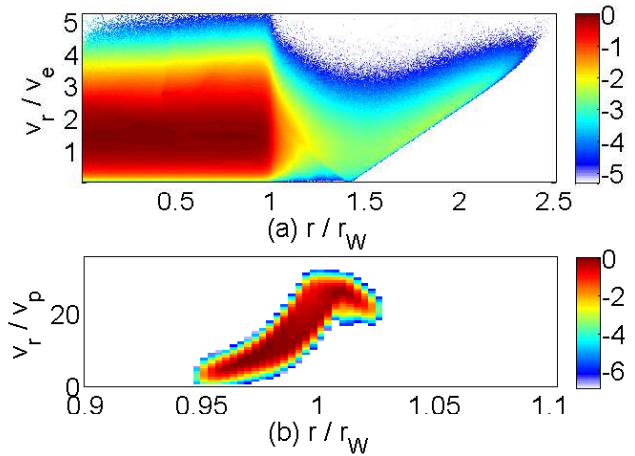}%
\caption{(Color online) The 10-logarithmic phase space densities $f_r (r,v_r)$ 
of the electrons (a) and $f_r (r,v_r)$ of the protons (b). The densities are 
normalized to their initial value. The simulation time is $t=27$ (Enhanced online).}
\label{f4}
\end{figure}

The phase space density distributions $f_e (r,v_r)$ and $f_p (r,v_r)$ of electrons and protons, 
which are functions of the radius $r={(x^2+y^2)}^{1/2}$ and of the radial velocity component 
$v_r={(v_x^2+v_y^2)}^{1/2}$, are displayed in Fig. \ref{f4}. They are derived as follows. The 
circular symmetry of the cloud and the energy exchange between electrons (s=e) and the mobile
protons (s=p), which is at least initially limited to the x,y-plane and a function of the radius, 
imply that bi-Maxwellian distributions $\hat{f}_s(r,v_r,v_z)$ will develop for both species. We 
neglect the $v_z$ direction and define the average over the azimuthal angle $\rho$ 
\begin{equation}
f_s(r,v_r) = {(2\pi r)}^{-1}\int_{\rho=0}^{2\pi} \hat{f}_s(r,v_r) rd\rho,
\end{equation}
which greatly improves the visualized dynamical range of the phase space densities. We normalize 
$f_s(r,v_r)$ by its maximum value at $t=0$. The resulting phase space densities are thus constant 
as a function of $r$ at $t=0$ within the radial interval the mobile species occupy. 

The protons have been accelerated to several times $v_p$ at $t=27$ and the proton mean speed 
increases with the radius. The protons have, however, not moved far beyond $r_W$ during this
short time. The fastest electrons have reached a radius $r\approx 2.3 r_W$. Their density 
profile shows a straight line from $r\approx 1.5r_W$ and $v_r \approx 0$ to $r\approx 2.3 r_W$ 
and $v_r\approx 4$. These are the electrons that escaped from the cloud before the ambipolar 
electric field has fully developed. The linear density profile simply reflects that faster 
electrons have propagated to larger radii during $t=27$. The ambipolar electric field, which 
has been built up after the first electrons escaped into the vacuum (See Fig. \ref{f2}(c)), 
affects the electrons at lower $r$, and we can observe a decrease of the peak electron speed 
as we go from $r\approx 2.3 r_W$ to $r\approx 1.5 r_W$. The electrons lose kinetic energy as 
they overcome the electrostatic ambipolar potential. This potential is also responsible for the 
drastic drop of the electron number density at $r\approx r_W$. It is evident from Fig. \ref{f4} 
that the electron charge for $r>1.03r_W$ is not compensated by a proton charge. The electric 
field at $r>1.03 r_W$ in Fig. \ref{f3} is thus sustained by the electron sheath. 

\begin{figure}\suppressfloats
\includegraphics[width=\columnwidth]{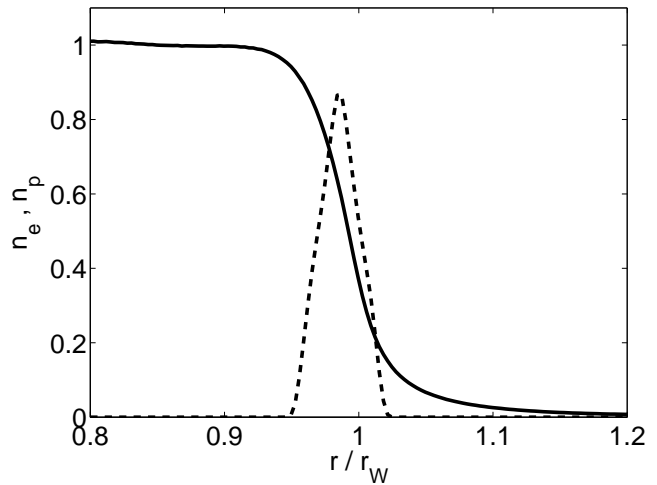}%
\caption{The electron density $n_e (r)$ and the proton density $n_p (r)$ (dashed
curve) sampled at the time $t=27$ and normalized to the respective initial density values.}
\label{f5}
\end{figure}

The density distributions of the electrons and the mobile protons are shown in Fig. \ref{f5}. 
They are obtained from the integration of the phase space densities in Fig. \ref{f4} over $v_r$. They are normalized to their initial value. The 
density distributions confirm that an electron sheath has formed outside the expanding proton 
cloud. 
The proton density has spread from the initial interval $0.95 < r/r_W < 1$ to $0.95 < r/r_W < 
1.03$. Note that we have not plotted here the charge density inside the proton ring distribution,
The density equals 1 in our normalization for $r < 0.95r_W$ and it ensures an overall charge 
neutrality of the plasma. The density distributions of the mobile species demonstrate that the 
interval $0.98 < r/r_W < 1.02$ is positively charged.  

\begin{figure}\suppressfloats
\includegraphics[width=\columnwidth]{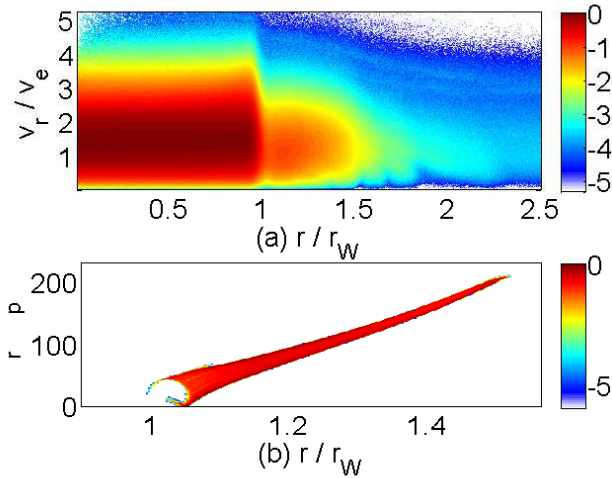}%
\caption{(Color online) The 10-logarithmic phase space densities $f_r (r,v_r)$ of 
the electrons (a) and $f_r (r,v_r)$ of the protons (b). The densities are normalized 
to their initial value. The simulation time is $t=500$.}\label{f6}
\end{figure}

The electron and proton phase space density distributions $f_e(r,v_r)$ and $f_p(r,v_r)$ at the 
time $t=500$ are shown in Fig. \ref{f6}. 
The majority of the electrons in Fig. \ref{f6}(a) is confined by the immobile positive charge 
background in the interval $r<0.95r_W$. Their phase space density and the characteristic 
electron speed decrease drastically at $r\approx r_W$ but remain relatively high up to 
$r\approx 1.5r_W$. The phase space density decreases by another two orders of magnitude as we 
go to even larger radii. The radial interval $1 < r / r_W < 1.5$ with the elevated electron 
phase space density and mean speed coincides with the radial interval that is occupied by the 
distribution of mobile protons. An almost closed circular proton phase space structure is present 
at $r\approx 1.03 r_W$ in Fig. \ref{f6}(b), which is similar to an ion phase space hole, and it 
must be related to a local excess of negative charge. The profile of the proton phase space 
density distribution in Fig. \ref{f6}(b) increases linearly with the radius for $r>1.05 r_W$, 
which is characteristic for a rarefaction wave. The protons reach a peak speed $\approx 200 v_p$, 
which is comparable to $v_e / 10$ and about the same as that in Ref. \cite{WireSim}. Only a 
small fraction of the protons reaches this speed, which limits the loss of electron thermal 
energy at this time (See Fig. \ref{f2}). The low electron phase space density for $r>2r_W$ 
implies that all electron processes close to the boundary are slow and do not carry much energy, 
which justifies our choice of periodic boundary conditions.

\begin{figure}\suppressfloats
\includegraphics[width=\columnwidth]{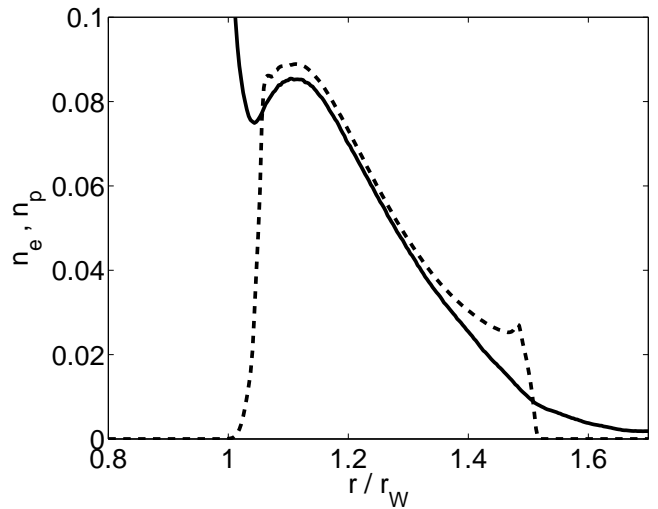}%
\caption{The electron density $n_e (r)$ and the proton density $n_p (r)$ (dashed curve) sampled 
at the time $t=500$ and normalized to the respective initial density values.}\label{f7}
\end{figure}

The densities of the mobile particle species at $t=500$ are shown in Fig. \ref{f7}.
The electron density decreases by an order of magnitude close to the boundary of the immobile
positive charge background at $r = 0.95r_W$. It increases again for $r> r_W$ and reaches a local 
maximum at around $r\approx 1.15r_W$, close to the peak of the mobile proton's density. Both
densities decrease gradually beyond this radius. The proton density falls off steeply at its
front at $r\approx 1.5 r_W$ (See also Fig. \ref{f6}(b)). 

\begin{figure}\suppressfloats
\includegraphics[width=\columnwidth]{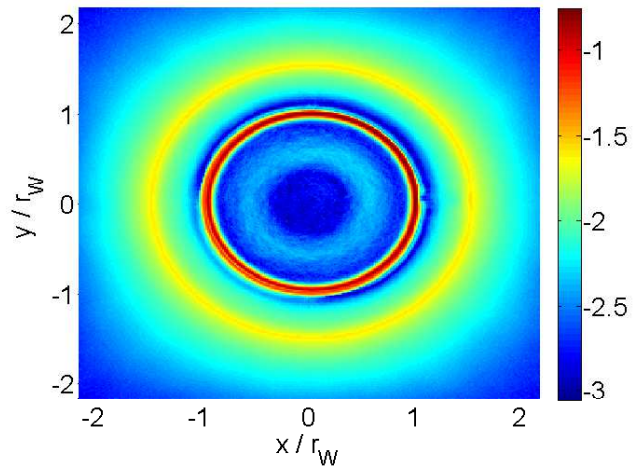}%
\caption{(Color online) The 10-logarithmic modulus of the electric field $|E_p(x,y)|$ 
sampled at the time $t=500$.}\label{f8}
\end{figure}

Figure \ref{f8} shows the in-plane electric field distribution at $t=500$.  
The electric field modulus shows a more complex pattern than at the earlier time. The electric
field peaks at $r\approx r_W$ and it is sustained by the electron density gradient close to
the boundary of the immobile positively charged background at $r=0.95 r_W$. The electric field
modulus goes through a minimum at a slightly larger radius. The reason for this radial oscillation 
is that we have two regions with an excess of positive charge, which are separated by a radial 
interval with a negative excess charge at $r\approx r_W$ (See Fig. \ref{f7}). An electric field 
with a significant amplitude modulus that extends over a large radial interval is observed at 
$1.1 < r/r_W < 2$. The electric field modulus in this band peaks at $r\approx 1.5 r_W$, which 
coincides with the front of the proton distribution in Fig. \ref{f7}. It thus corresponds to the 
ambipolar electric field driven by this density gradient, and it decreases for larger radii where 
we only find electrons. The electric field modulus reaches its minimum at $r\approx 2 r_W$. 

\begin{figure}\suppressfloats
\includegraphics[width=\columnwidth]{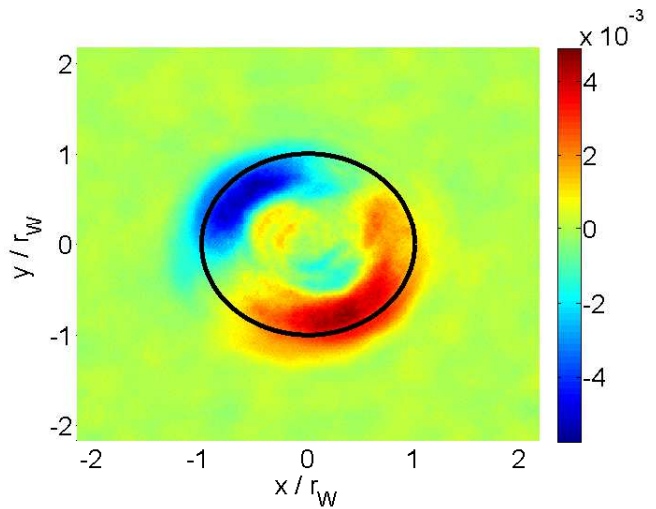}%
\caption{(Color online) The magnetic field amplitude $B_z(x,y)$ sampled at the time $t=500$. 
Overplotted is a circle of radius $r=r_W$ (Enhanced online).}\label{f9}
\end{figure}

Figure \ref{f9} displays the distribution of $B_z$ at $t=500$. We observe a strong localized 
magnetic field structure. Its amplitude peaks at $r\approx 0.8r_W$ and it decreases as we go 
to $r\approx r_W$ and beyond. The amplitude oscillates once in the azimuthal 
direction and the associated wavelength $\lambda_\rho \approx 2\pi r_W$ is thus much larger 
than the one in the radial direction, which we can estimate as follows. We find a magnetic 
field maximum at $x \approx -0.3 r_W$ and $y\approx 0.3r_W$ and a minimum at $x\approx -0.6r_W$ and $y \approx 0.6 r_W$. The 
wavelength of the radial oscillation is thus $\lambda_r \approx 0.5r_W$. The magnetic amplitude 
outside this radial interval is at noise levels. The magnetic noise is distributed over the 
entire simulation box, while the magnetic structure is localized in a small radial interval. 
This explains why we do not observe a more pronounced growth of the magnetic field energy in 
Fig. \ref{f2}(d).

It is instructive to compare the radial interval, in which the magnetic field grows, with the 
one that shows a thermal anisotropy. We determine for this purpose the thermal energy densities 
of the electrons in the radial and azimuthal directions. We consider only the in-plane component 
of the speed $\mathbf{v}_{p,j} = {(v_x,v_y)}_j$ of the $j^{th}$ computational electron. The
radial component of the thermal energy $K_{r,j} = v_{r,j}^2$ is computed from the projection 
$v_{r,j} = \mathbf{v}_{p,j} \cdot \mathbf{r}_j/r_j$, where $\mathbf{r}_j$ is the position vector
of the electron in circular coordinates, and $K_{\rho,j} = v_{p,j}^2-K_{r,j}$. The partial thermal 
energies and the anisotropy $A$ are then obtained from the summations
\begin{eqnarray}
K_r (i\delta_r) = \sum_{j=1}^{N_e} K_{r,j} \delta_{i,j} \\
K_\rho (i\delta_r) = \sum_{j=1}^{N_e} K_{\rho,j} \delta_{i,j} \\
A = K_r / K_\rho, 
\end{eqnarray}
where $\delta_{i,j}=1$, if $r_j$ falls into the $i^{th}$ radial bin of width $\delta_r$, and zero 
otherwise. We thus obtain a histogram $A(i)$ of the radial distribution of the thermal anisotropy.
The anisotropy $A(r)$ is compared with the magnetic energy 
\begin{equation}
P_{Bz}(n\delta_r)=(\Delta_x^3 / 2\mu_0)
\sum_{i,j=1}^{N_g} B_z^2(i-N_g/2,j-N_g/2)\delta_{i.j,n}, 
\end{equation}
where $\delta_{i,j,n}=1$ if $I(i^2 + j^2) = n^2$, with $I$ being a round-off operation and $\Delta_x = \delta_r$. This azimuthal 
integration, rather than the azimuthal average, emphasizes magnetic fields at larger radii. 

\begin{figure}\suppressfloats
\includegraphics[width=\columnwidth]{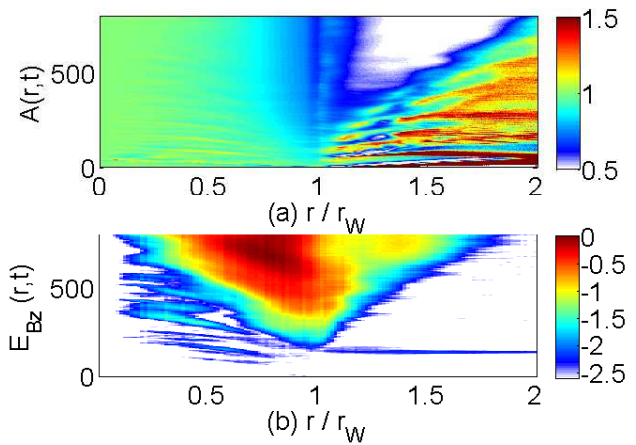}%
\caption{(Color online) The ratio $A$ between the mean radial energy and the mean perpendicular 
energy of electrons is shown on a linear color scale in (a). The spatio-temporal evolution of the 
magnetic energy density is shown on a 10-logarithmic color scale in (b).}\label{f10}
\end{figure}

Figure \ref{f10} demonstrates that the magnetic field starts to grow at $t\approx 200$, when an 
anisotropy $A(r)<1$ has formed that is sufficiently strong and wide. 
The magnetic field grows initially in the interval $0.7 < r/r_W < 1.1$, but it expands later on 
in both radial directions. Its front reaches $r\approx 1.5 r_W$ at $t=500$, which coincides with 
the tip of the proton distribution in Fig. \ref{f7}. It is thus confined to within the rarefaction 
wave. The radial interval where $P_{BZ} \approx 1$ is initially stationary, but the magnetic 
field distribution changes after $t \approx 500$. What appears to be a sidelobe develops at $r 
\approx 1.4r_W$. This change takes place on a few tens of $\omega_p^{-1}$. The plasma frequency 
in the rarefaction wave is about $\omega_p / 4$ at $t=500$ (See Fig. \ref{f7}) and the growth 
time of the sidelobe is thus faster than that expected from any instability. 
Indeed, the on-line enhancement of Fig. \ref{f9} shows that the magnetic $B_z$-field leaks out 
from within the cloud into the rarefaction wave. 

\begin{figure}\suppressfloats
\includegraphics[width=\columnwidth]{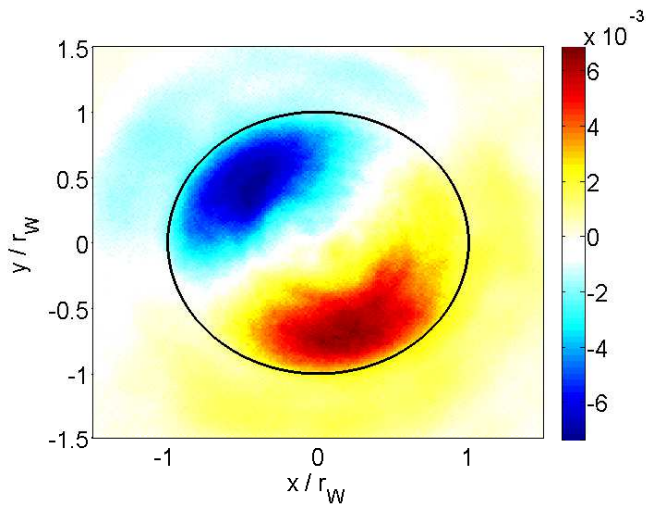}%
\caption{(Color online) The magnetic field amplitude $B_z(x,y)$ sampled at the time 
$t=800$. Overplotted is a circle of radius $r=r_W$.}\label{f11}
\end{figure}

Figure \ref{f11} shows its distribution at $t=800$.  
The extrema of $B_z$ are located within $r=r_W$ and the amplitude of $B_z$ is constant as a 
function of $r$ for $1<r/r_W<1.5$. The magnetic energy has been confined to within $r\approx r_W$ 
until $t\approx 600$ and it forms a TM mode until then (See Fig. \ref{f9}). The rapid expansion 
of the magnetic energy after $t\approx 600$ implies that this TM mode suddenly expands. 
 
We can understand the circular plasma cloud as the cross section of a cylindrical waveguide with
an axis that is aligned with $z$. The expansion of the mobile protons imply that the radius of
the cross-section of this waveguide increases in time. It is well-known that changes in the
radius of a waveguide induce a coupling of TM and TE modes \cite{Waveguide}. The magnetic field
of a TE mode would be oriented in the simulation plane. 

\begin{figure}\suppressfloats
\includegraphics[width=\columnwidth]{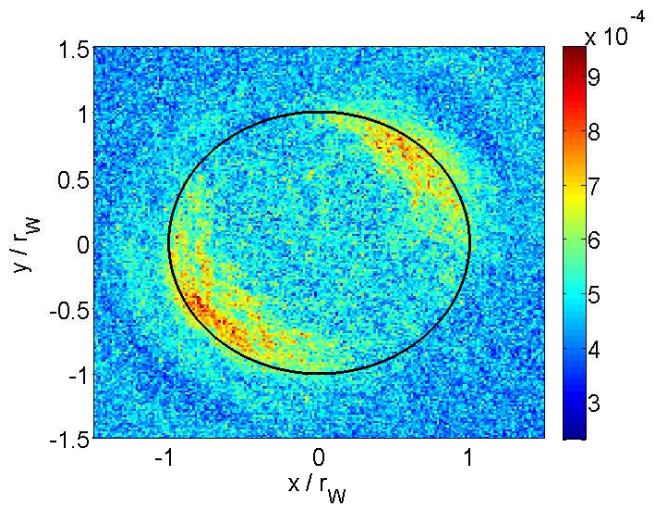}%
\caption{(Color online) The magnetic field amplitude $|B_p(x,y)|$ sampled at the time 
$t=800$. Overplotted is a circle of radius $r=r_W$.}\label{f12}
\end{figure}

Figure \ref{f12} evidences that a TE wave is indeed present at $t=800$. 
A structure is visible in the distribution of the in-plane magnetic field $B_p = {( B_x^2 + 
B_y^2 )}^{1/2}$. It is located in the same radial interval as the magnetic field of the TM mode. 
The magnetic field patterns in $B_z$ and $B_p$ have the same azimuthal wave number and both are 
phase-shifted by $90^\circ$ relative to each other. The magnetic energy of the TE mode grows after 
$t=800$ to values exceeding that of the TM wave (not shown). We do not discuss the evolution for 
$t>800$. The TE wave expands out to the boundaries shortly after this time, which results in 
finite box effects, while the TM wave remains confined by the rarefaction wave as discussed previously \cite{WireSim}. 

The different behaviour of the TM and TE modes can be explained by the different plasma response to their electric fields in our 2D box geometry. An electric field orthogonal to the simulation box plane can not result in charge 
density modulations, since we do not resolve the z-direction. It thus only affects the current 
distribution. An in-plane electric field does, however, modulate also the charge density. The 
dilute electron plasma between the front of the rarefaction wave and the boundaries can not 
support strong charge density waves, but it can easily support the large currents from the TE wave
and the latter can expand more easily. 
\section{Discussion}

We have modeled here with a 2D PIC simulation the expansion of a circular plasma cloud into a
vacuum, which has been driven by the thermal pressure of the electrons. It is a follow-up study 
of a previous simulation experiment. It aimed at explaining the growth of magnetic fields in
the rarefaction wave, which is generated by the ablation of a wire by a laser pulse \cite{WireExp}. 
It considered the expansion of a circular plasma cloud, which consisted of spatially uniform hot
(32 keV) electrons and cool (10 eV) protons. Here we have confined the mobile protons to the 
border of the plasma cloud. This hollow ring distribution is a more accurate approximation of 
the experimental conditions. The rarefaction wave observed in Ref. \cite{WireExp} contains 
primarily the light ions from the surface impurities, which have been ionized by the strong 
surface electric field and current \cite{Surface}. However, computational constraints require 
us to represent here the electrons as a hot (32 keV) species that is uniformly distributed over 
the entire plasma cloud. The electrons in the experiment reach MeV temperatures, but they are 
confined to the wire's surface. Choosing cooler electrons reduces the difference between the 
electron and proton Debye lengths, which is computationally efficient, while it ensures that 
the thermal energy that drives the expansion is comparable in simulation and experiment.

Our results are as follows. The ambipolar electric field driven by the electron's thermal 
expansion results in the formation of a rarefaction wave. The fastest protons reach a speed 
that is comparable to about a tenth of the electron thermal speed, which equals the value 
observed in Ref. \cite{WireSim}. The proton acceleration is thus not affected by the choice 
of the initial proton distribution. However, the density of the rarefaction wave is limited 
by the number of available mobile protons. The spatially uniform proton distribution in Ref. 
\cite{WireSim} provided a continuous feed of mobile protons, while the number of mobile protons 
we introduce here is limited. The plasma density in the rarefaction wave we observe here is
thus lower than that in Ref. \cite{WireSim}. The therefrom resulting lower plasma frequency 
implies a slowdown of the instabilities in the rarefaction wave. 

Although the thermal anisotropy in the electron distribution here and in Ref. \cite{WireSim}
has been comparable, the Weibel-type instability could not develop here during the simulation 
time. The instability developed instead in the dense core of the plasma cloud and the magnetic 
field diffused out into the rarefaction wave. This magnetic diffusion is not compatible with 
the magnetic field structures observed in Ref. \cite{WireExp}, which can be explained better 
by an instability in the rarefaction wave. Remarkably the more realistic proton distribution 
we use here yields results that are less representative for the experimental ones than those 
of the simulation in Ref. \cite{WireExp}. This is probably a consequence of the spatially 
uniform electron distribution. The MeV electrons form a thin layer on the wire's surface in 
the experiment and the magnetic instability can not form in it. It can only form in the 
rarefaction wave, which maintains a significant thermal anisotropy over a large radial interval. 
Future studies thus have to confine the electrons to a smaller radial interval.

The present simulation sheds light on the mechanism by which the in-plane magnetic fields
grew in Ref. \cite{WireSim}. The plasma in the radial interval $r<0.95r_W$ with the immobile 
positive charge background is equivalent to a cylindrical waveguide and the thermal anisotropy
of the electrons results in the growth of a TM wave inside the wave guide. The plasma of the
dilute rarefaction wave is a perturbation of the waveguide's cross section. A varying radius
of a waveguide can couple TM and TE modes \cite{Waveguide}. Here this coupling results in
the growth of in-plane magnetic fields. A TE mode can probably not be driven by a plasma
instability. The electromagnetic forces and the plasma flow are confined to within the 
simulation plane. No current can thus develop in the orthogonal direction. Orthogonal plasma
currents are, however, needed to maintain the in-plane magnetic field. A wave coupling between 
a TM and a TE wave drives orthogonal electric fields, which can accelerate electrons in this 
direction. Future work will address this wave coupling with a larger simulation box. The
evolution of the TE wave can then be examined for a longer time without finite box effects.

{\bf Acknowledgements:} MED thanks Vetenskapsr\aa det (Grant 2010-4063), GS thanks the 
Leverhulme foundation (Grant ) and MB thanks EPSRC (Grant) for financial support. Computer 
time and support has been provided by the HPC2N computer center in Ume\aa , Sweden.

\end{document}